\documentclass[12pt]{article}

\usepackage{graphicx,amsmath}
\usepackage{authblk}

\begin{document}

%\title{}
%\author{}
\date{}
%\maketitle
%
%

\title{Electronic structure of Gold, Aluminum and Gallium Superatom Complexes}
\author[1]{O. Lopez-Acevedo\thanks{lopez@phys.jyu.fi}}
\author[1]{P. A. Clayborne}
\author[1,2]{H. H\"akkinen}
\affil[1]{Nanoscience Center, Department of Chemistry, University of Jyv\"askyl\"a, 40014 Jyv\"askyl\"a, Finland}
\affil[2]{Nanoscience Center, Department of Physics, University of Jyv\"askyl\"a, 40014 Jyv\"askyl\"a, Finland}

%\affiliation{Nanoscience Center, Department of Chemistry, University of Jyv\"askyl\"a, 40014 Jyv\"askyl\"a, Finland}
%\email{lopez@phys.jyu.fi}
%\author{P. A. Clayborne}
%\affiliation{Nanoscience Center, Department of Chemistry, University of Jyv\"askyl\"a, 40014 Jyv\"askyl\"a, Finland}
%\author{H. H\"akkinen}
%\affiliation{Nanoscience Center, Department of Chemistry, University of Jyv\"askyl\"a, 40014 Jyv\"askyl\"a, Finland}
%\affiliation{Nanoscience Center, Department of Physics, University of Jyv\"askyl\"a, 40014 Jyv\"askyl\"a, Finland}
%\date{\today} % delete this line to display the current date
\maketitle

\begin{abstract}
Using ab-initio computational techniques on crystal determined clusters, we report on the similarities and differences of  Al$_{50}$(C$_5$(CH$_3)_5)_{12}$, Ga$_{23}$(N(Si(CH$_3)_{3}$)$_{2}$)$_{11}$, and Au$_{102}$(SC$_7$O$_2$H$_5$)$_{44}$ ligand-protected clusters. Each of the ligand-protected clusters in this study show the similar stable character which can be described via a electronic shell model. We show here that the same type of analysis leads consistently to derive a superatomic electronic counting rule, independently of the metal and ligand compositions. One can define the cluster core as the set of atoms where delocalized single-angular-momentum-character orbitals have hight weight using a combination of Bader analysis and the evaluation of Khon-Sham orbitals. Subsequently one can derive the nature of the ligand-core interaction.  These results yield further insight into the superatom analogy for the class of ligand-protected metal clusters.
\end{abstract}

%\keywords{Electron shell model, superatom model, metal cluster, ligand-protected cluster}
%\pacs{73.22.-f,36.40.-c,61.46.-w}

\maketitle
%\tableofcontents
\section{Introduction}
Protected gold clusters have been synthesized in several sizes and compositions. Due to their intrinsic stability and potential applications in nanotechnology, they have received broad interest in past decades. For example, size dependent optical, electrochemical and catalytical properties have been experimentally determined \cite{CW98,SW00,NT05,CT08,B94,H08,SM09,J10}. In addition, these clusters can be expected to be organized as building blocks of materials with new interesting properties. Recent breakthroughs such as determining the crystal structure of ligand-protected clusters\cite{JK07,HM08,ZJ08,FZ08,QJ10} , has led to further investigations of their electronic structure via ab-initio simulations . As a result, the electronic structure and derived properties of ligand-protected gold clusters can be modeled via a modified electronic shell model, termed superatom model \cite{WH08,AH08,LH10,LA10}. 

Electronic shell models have been successfully used to understand and predict properties of bare metallic clusters\cite{KC84,C85,H93}.  One possible spherical shell model is a 3D harmonic oscillator with an anharmonic, angular momentum dependent, term.  From this model several properties are derived, most notably the high stability of some clusters. For a given composition, the model predicts stable clusters with large HOMO-LUMO gaps corresponding to the closing of an electronic shell level. 
The order of the shells and the magic numbers in this spherical shell model depends then on the anharmonic parameter. For an anharmonic parameter in the intermediate region the order of the shells is:
$$1S^2 ~ 1P^6 ~ 1D^{10} ~ 2S^2 ~ 1F^{14}~  2P^6 ~ 1G^{18} ~ 2D^{10} ~ 3S^2 ~ 1H^{22}  ~2F^{14}~  3P^6 ~ 1I^{26} ~ 2G^{18}$$
predicting magic numbers at 58 and 138 electrons, among others, corresponding to the closing of the $1G$ and $1I$ shell respectively ($U=0.03$, equation $A_1$\cite{H93}).

Aluminum and gallium metalloid clusters (clusters containing more metal-metal bonds than metal-ligand bonds) have also been characterized experimentally and theoretically in an effort to understand how properties evolve from clusters to bulk and their stability has been explained through various models \cite{W72,M84,S10}. However, reports on bare aluminum clusters have shown the most stable species have superatomic character with a magic number of electrons, which adheres to the electronic shell model \cite{LC89,CK09}. For example, the Al$_{13}^{-}$ cluster is resistive to O$_2$ etching with 40 electrons (magic number), while its neutral counterpart (Al$_{13}$) is defined as a superhalogen based on its similar electron affinity to halogens on the periodic table \cite{LC89,BK04,BK05,RC06,RC07}. This result, along with the studies on ligand-protected gold clusters, suggest the electronic shell model could describe metalloid clusters composed of aluminum and gallium. Recently our group has analyzed aluminum metalloids and successfully illustrated a superatom electronic structure exists which relates to a cluster's overall stability\cite{CH10}. For few gallium metalloid clusters, it has been predicted the electronic shell model may be successful; however to our knowledge no theoretical investigations into the electronic structure have been performed.

In order to derive an electronic shell structure for nearly spherical ligand-protected clusters, we currently use an orbital projection on spherical harmonics integrated on the cluster region\cite{WH08}. 
The orbitals can be labelled with a given angular momentum using the coefficients and one can determine the order of electronic shells in the cluster system. It is necessary however to take the role of the ligand layer into account to determine the expected number of delocalized orbitals participating in the shell structure. In the superatom model, one takes the ligand-core interaction into account.  In a ligand-protected superatom cluster the core atoms participate collectively to give rise to a cluster size delocalized orbitals; while the protective units participate only in localized or interface bonding states. The protective ligands can behave by either depleting or donating electrons (or remaining neutral) to the superatom electronic structure and subsequently will give rise to different electron counts i.e-magic numbers. The core-ligand interaction is not only important from a theoretical perspective, but proves to have implications in experimental observations as well. \cite{SH10,LA10}  

Completing the characterization of the clusters as ligand-protected superatoms requires an understanding of not only the electronic structure, but the ligand-core interaction as well. We show using ab-initio computational techniques, three different ligand-protected clusters (Al$_{50}$Cp$^*_{12}$ Cp*=C$_5$(CH$_3)_5$ and Ga$_{23}$L$_{11}$, L = N(Si(CH$_3$)$_3)_2$ and Au$_{102}$(SR)$_{44}$, R=C$_7$O$_2$H$_5$) can be fully characterized as superatoms. Projection of the Kohn-Sham (KS) orbitals on a local atomic basis is used to determine which atomic layers contribute to form the delocalized orbitals and are part of the cluster core.  Using Bader analysis we find one can characterize the local atomic electronic structure and gain information into the nature and description of the ligand-protecting shell for each cluster.  Based on these analyses of the $M_N[X_Y]^z$ clusters, one can predict the number of delocalized electrons using the equation:
\begin{equation}\label{counting} 
n_e^* = NV_M-YV_X-z
\end{equation}
where N is the number of atoms in the core with valence $V_M$, Y is the number of protective units depleting $V_X$ electrons each and the cluster has an overall charge of z, that contribute to the superatomic orbitals in the superatom model.  Finally, we compare the similarities and differences of the three superatom complexes and their expected properties.

\section{Computational methods}
The computations were done using the GPAW code, which performs calculations based on Density Functional Theory \cite{KS65,GPAW}. The code is a grid based implementation of the projector-augmented wave method (PAW). Furthermore a frozen core approximation is used. H(1s), C(2s2p), Al(3s3p), S(3s3p), O(2s2p), Au(5d1s), N(2s,2p), Si(3s,3p), Ga(4s,4p) are treated in the valence. The exchange-correlation functional used in all calculations is PBE \cite{PE96}. Relaxation of the system is performed until the forces around all atoms are below 0.05 eV/\AA. The crystal structures for the Au$_{102}$(SR)$_{44}$, Al$_{50}$Cp$^*_{12}$  and Ga$_{23}$L$_{11}$ were obtained from the experimentally reported structures through the CCDC database\cite{HS07,VS05,JK07}. From the crystal structure the coordinates of a single cluster were isolated and were allowed "in vacuum" to optimize without constraints.

Atomic charge state (extra or missing local charge with respect to the atomic number) is determined using a Bader type of analysis \cite{B90,TH09}. 
Projecting the all electron partial waves (i.e. the wave functions of the isolated atoms) into molecular orbitals (PLDOS) is done within the PAW formalism following \cite{GPAW}.

The superatomic analysis is done by a projection of the KS orbitals on spherical harmonics \cite{WH08}. For simplicity the origin is chosen in the center of mass of the cluster. For a given core radius $R_0$, the angular momentum weight $c_l$ associated to the KS orbital $\psi$ is defined using:
\begin{eqnarray} \label{YLM}
c_l = \sum_{m=-l}^l {f_l^m},\\ 
f^m_l =   \int_0^{R_0} \left| \int_{\Omega} {Y_l^m}^*(\theta,\varphi) \psi(r,\theta,\varphi)  d\Omega \right|^2 r^2 dr,
\end{eqnarray}
where $Y_l^m(\theta,\varphi)$ is a spherical harmonic function with degree $l$ and order $m$ and $d\Omega = \sin\theta d\varphi d\theta$.
\section{Results and discussion}

\subsection{Protected gold clusters Au$_{102}$(SR)$_{44}$}
\begin{figure}
\centering
\includegraphics{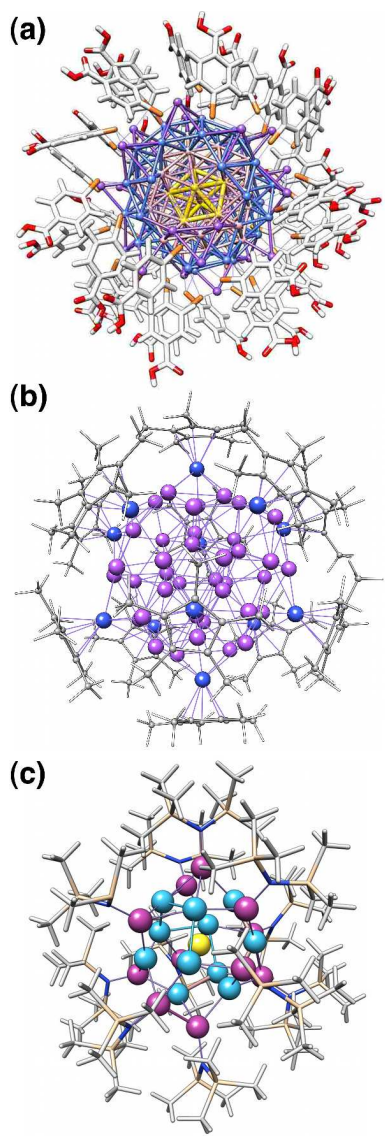}
\caption{Radial atomic decomposition of the ligand protected clusters considered. Metal atoms are represented with spheres and all others with sticks. Metal atoms in the same radial layer share the same color. Some atoms have been removed to enhance the view of the core. (a) Protected gold cluster Au$_{102}$(SC$_7$O$_2$H$_5$)$_{44}$: $r_1$, $r_2$, $r_3$ and $r_4$ are colored in yellow, pink, blue and violet respectively.  (b) Protected aluminum cluster Al$_{50}$(C$_5$(CH$_3)_5)_{12}$:  $r_1$ and $r_2$ are colored in violet and  $r_3$ in blue. (c) Protected gallium cluster Ga$_{23}$(N(Si(CH$_3)_{3}$)$_{2}$)$_{11}$:  $r_1$, $r_2$, $r_3$ are colored in yellow blue and violet respectively. }
\label{structure}
\end{figure}

\begin{figure}
\centering
\includegraphics[scale=0.6]{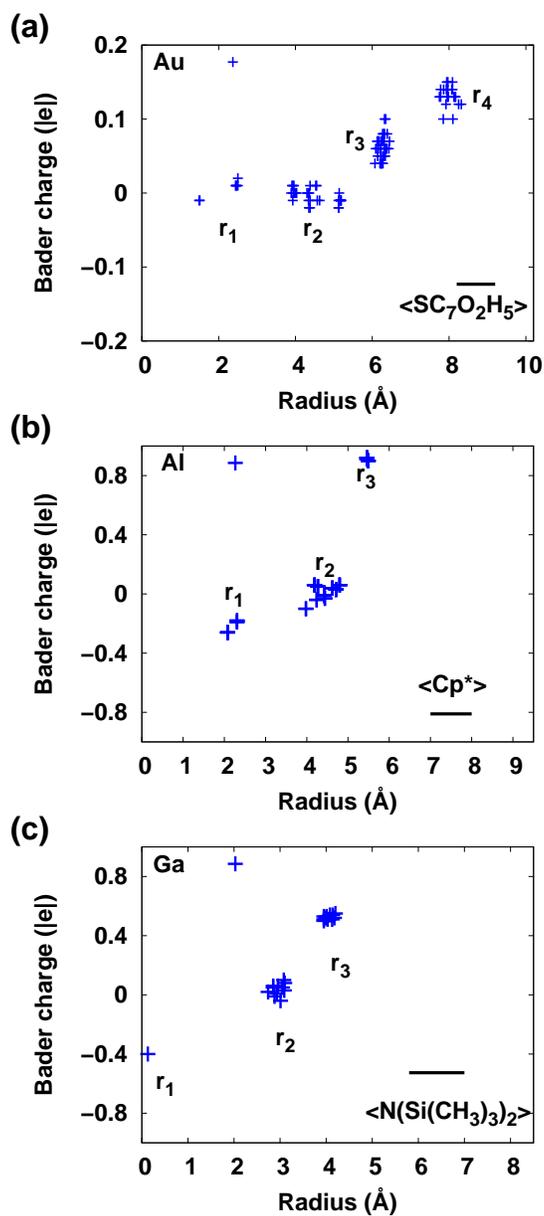}
\caption{Bader charge as a function of the radial position of the atoms. Negative charge indicates excess with respect to the neutral atom charge. The origin of the coordinates is chosen in the center of mass of the cluster. Metal atoms are all included individually. The rest of the charge is indicated as a per ligand average. (a) Protected gold cluster Au$_{102}$(SC$_7$O$_2$H$_5$)$_{44}$  (b) Protected aluminum cluster Al$_{50}$(C$_5$(CH$_3)_5)_{12}$ (c) Protected gallium cluster Ga$_{23}$(N(Si(CH$_3)_{3}$)$_{2}$)$_{11}$.}
\label{bader}
\end{figure}

\begin{figure}
\centering
\includegraphics[scale=0.6]{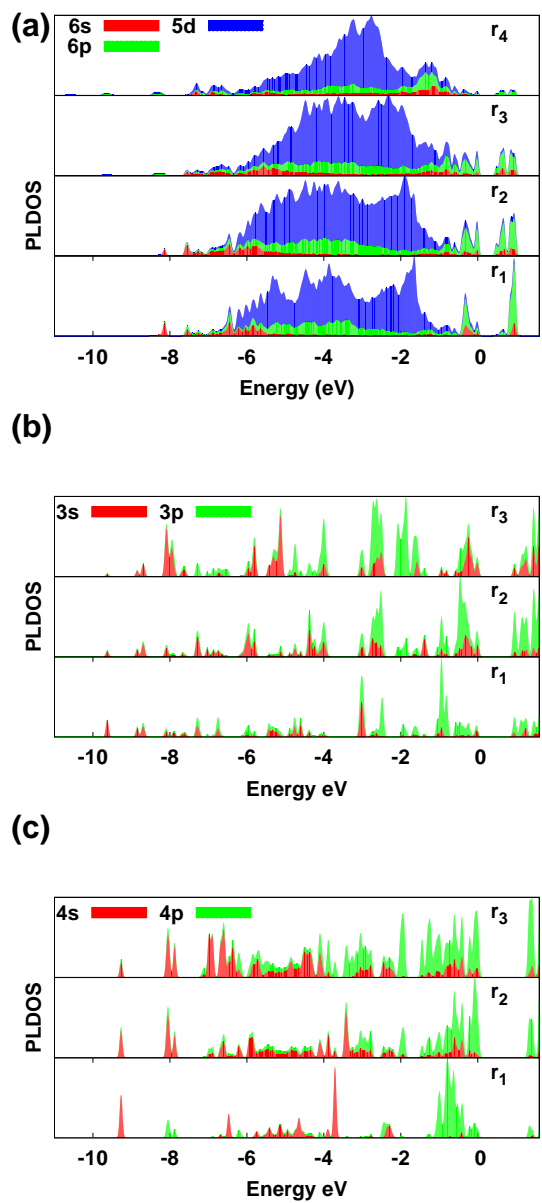}
\caption{Projected Local Density of States using atomic basis. The projection is integrated in the radial layers composed by metal atoms. (a) Protected gold cluster Au$_{102}$(SC$_7$O$_2$H$_5$)$_{44}$  (b) Protected aluminum cluster Al$_{50}$(C$_5$(CH$_3)_5)_{12}$ (c) Protected gallium cluster Ga$_{23}$(N(Si(CH$_3)_{3}$)$_{2}$)$_{11}$. }
\label{dos}
\end{figure}

First it is important to see in the protected gold cluster that the gold atoms can be divided in two sets: the core and the ligand set \cite{HG08}. As this result has been derived previously and generalized to other protected gold clusters we include here the discussion for completeness and as an ilustration of the method that will be utilized in the aluminum and gallium case.
The  Au$_{102}$(SR)$_{44}$ cluster can be viewed in subsequent layers, Figure \ref{structure} (a). We find from the analysis that the atoms in the first three layers $r_1$ to $r_3$ have each a very small average Bader charge, 0.01, 0.00 and 0.06 $\left | e \right |$ respectively in Figure \ref{bader} (a). The atoms in the outermost layer $r_4$ show a small but distinct positive 0.13 $\left | e \right |$ mean charge. The projection on an atomic basis for the inner gold atoms (layers $r_1$ to $r_3$) shows a s-p hybrid band with high weights around higest occupied molecular orbital (HOMO) in the PLDOS Figure \ref{dos} (a). In contrast there are almost no weight on the orbitals around HOMO in the external layer $r_4$. Those KS orbitals also have high s-p atomic local component, are delocalized superatom orbitals. The superatomic projection shows a change of angular momentum G to H and corresponds to a clear gap of 0.48 between the HOMO and the lowest unoccupied molecular orbital (LUMO) \cite{WH08}, see Figure \ref{ylm} (a).
As a consequence, the superatom core is composed of the first three shells $r_1$ to $r_3$ only; the core being covered by 21 units (Au(SR)$_2)_{19}$ and (Au$_2$(SR)$_3)_{2}$ including the gold atoms belonging to the shell $r_4$. One can rewrite then the cluster formula, making explicit the core-ligand structure as follows: Au$_{79}$[(Au(SR)$_2)_{19}$(Au$_2$(SR)$_3)_{2}]$ implying 79 gold atoms contribute with one electron (6s) to the delocalized super atom counting but 21 of those are depleted by the protective units. This gives there are 58 superatom electrons, corresponding to a cluster with electronic closed shell configuration in a spherical potential. 

\subsection{Protected aluminum Al$_{50}$(Cp*)$_{12}$}
The aluminum metalloid cluster can be view geometrically into three distinct shells of aluminum atoms, an inner Al$_8$ layer, encapsulated by 30 Al atoms with an exterior shell of twelve Al atoms Figure \ref{structure} (b). The exterior twelve aluminum atoms are bonded to twelve pentamethylcyclopentadienyl (Cp*) ligands.  It should be noted, whether the exterior layer should be considered part of the ligand or the core is currently a subject of discussion \cite{S10,SW11}. In order to understand if the outer twelve aluminum atoms are part of the ligand or not, we performed a similar ab-initio simulation analysis to determine the core and ligand sets as well as the electronic state of the atoms.

The twelve exterior Al atoms have an average Bader charge of 0.91 $\left | e \right |$ (Figure \ref{bader} (b)), while the innermost atoms has zero or negative charge. Thus, these 12 Al atoms donate 1 electron to the electron withdrawing ligand Cp*. The projection of KS orbitals on local atomic basis is described in Figure \ref{dos} (b) where it is shown that all shells, $r_1$ to $r_3$, contribute to the delocalized superatomic orbitals. The outermost shell $r_3$ near the HOMO level shows more s states than p per atom, which can be attributed to the loss of 1 p-electron per Al atom and corresponds to the Bader charge results. Further, the projection on spherical harmonics shows the Al$_{50}$(Cp*)$_{12}$ has an electronic structure corresponding to the one expected for a superatom cluster with a shell closing of 138 electrons as reported previously \cite{CH10} see Figure \ref{ylm}(b).  The HOMO-LUMO gap (0.94 eV) corresponds to the transition from the 1I shell to the 2G shell. The splitting of those 1I and 2G shells can be obtained within the spherical shell model with the anharmonic parameter indicated in the introduction. Further, when changing the radius parameter in the projection of spherical harmonics we confirm it is important to include the most outer aluminum atoms since they contribute to form the higher delocalized superatomic orbitals  (Figure 1 in Supporting Information). Thus, the combination of these results point to the characterization of this metalloid cluster as a ligand-protected superatom with ionic bonding to the Cp* ligands. The structural separation in core-ligand can be represented writing the chemical formula as Al$_{50}$[Cp*$_{12}$]. The 50 Al atoms of the core contribute each with three electrons  while each protective ligand Cp* deplete one electron giving a total number of superatom electrons of 138.

\subsection{Protected gallium Ga$_{23}$R$_{11}$}
The Ga$_{23}$(N(SiMe$_3)_2)_{11}$ metalloid cluster can also be viewed as a cluster containing multiple radial atomic layers as in the previous two cases \ref{structure} (c). The inner most layer consists of one gallium atom followed by two consecutive layers of eleven gallium atoms, with the outer eleven being surrounded by ligands.  Previously, the gallium metalloid cluster has been described as a Ga$_{12}$ core surrounded by 11 GaL units \cite{HS07}. We find the 11 Ga atoms on the exterior layer have an average Bader charge of 0.52 $\left | e \right |$, with the inner shell having a negative -0.40 and neutral 0.04 $\left | e \right |$ mean Bader charge (Figure \ref{bader} (c)).   The Bader charge value is indicative of polarized bonding, which should come as no surprise since the ligand is composed of nitrogen.  It is well-known that the nitrogen atom participate in polarized bonding.  From the projection on local atomic basis PLDOS (Figure \ref{dos} (c) ) it is obtained all gallium atoms are in the same electronic state and contribute to the delocalized superatom states around HOMO. The combination of Bader analysis and PLDOS allows us to conclude the gallium cluster should be described as a metallic core of 23 atoms protected by 11 electron depleting units. The projection of the KS orbitals on spherical harmonics reveals the ligand-protected gallium cluster does adhere to the superatom model with a closed 1G shell with 58 superatomic electrons (Figure \ref{ylm} (c)).  The cluster has a large gap of 1.34 eV, which is indicative of its stable nature as well. The derived core-ligand composition is thus Ga$_{23}$[R$_{11}$] with 23 gallium atom contributing with 3 electrons each and 11 protective units depleting one electron and thus corresponding to a counting of 58 superatom electrons.

\subsection{Comparison}
All three clusters considered in this study have a core that can be separated in concentric layers. Each layer being formed by metal atoms having not only the same radius but also the same local Bader charge.  Likewise, their electronic structure can be explained using the superatom model with the number of electrons contributing to the delocalized superatomic orbitals obtained using equation \ref{counting}.  Using this electron counting rule one obtains 58, 138 and 58 electrons for the Au$_{102}$(SR)$_{44}$, Al$_{50}$[Cp*$_{12}$ and Ga$_{23}$(N(SiMe$_3)_2)_{11}$ clusters, respectively.    
Each of the cluster are superatom complexes, but the projection on the local atomic basis shows differences in the layers that contribute to the delocalized superatomic orbitals.  For the gold cluster, the gold atoms in the protective exterior do not participate to form such orbitals.  In the cases of the metalloid superatom complexes their exterior metal atoms do contribute to form the superatomic orbitals. Further differences can be seen from the Bader analysis.  The charge distribution in the ligand-protected aluminum and gallium clusters proceeds from negative to neutral to positive from the center to outer shells. However, in the ligand-protected gold cluster  most inner two shells ($r_1$ and $r_2$) are neutral, with the third shell $r_3$ being only slightly positive and the exterior shell $r_4$ positive as well. Another difference is found when focusing on the ligand and exterior atomic layer interaction.  The exterior metal atoms in all three cases have very different Bader values with the aluminum value being the largest (0.91 $\left | e \right |$) and gold the smallest.  It is interesting that the gallium value is 0.52 $\left | e \right |$, yet still positive as for both aluminum and gold.  These differences point to the nature of the ligand-metal bonding in the superatom complexes.  It is well-known that the bonding for many ligand-protected gold clusters is covalent. We have shown one can have ionic and polarized covalent bonding in other superatom complexes as seen in the aluminum and gallium cases presented here. 

\begin{table}
   \centering
   %\topcaption{Table captions are better up top} % requires the topcapt package
   \begin{tabular}{l|l} % Column formatting, @{} suppresses leading/trailing space
     
  %    \multicolumn{2}{c}{Formulas} \\
%      \cmidrule(r){1-2} % Partial rule. (r) trims the line a little bit on the right; (l) & (lr) also possible
      $M_N [X_Y]$&$n_e =   NV_M-YV_X-z$\\
      \hline
       %\toprule
%      \midrule
        Au$_{79}$[(Au(SR)$_2)_{19}$(Au$_2$(SR)$_3)_{2}]$& $n_e = 79 - 21 = 58$\\ 
         Al$_{50}$[Cp*$_{12}$] & $n_e = 3 \times 50 - 12 = 138$\\
         Ga$_{23}$[R$_{11}$]& $n_e = 3 \times 23 - 11 = 58$
%      \bottomrule
   \end{tabular}
   \caption{Formulas and derived superatomic counting rule}
   \label{summary}
\end{table}
\section{Conclusion}
We have investigated three distinct ligand-protected clusters composed of gold, aluminum and gallium. The electronic structure of these clusters can be described via the superatom model and have magic numbers corresponding to 58, 138 and 58 electrons, respectively. One can also define the metallic core as the group of atoms which participate in the delocalized superatomic orbitals using the analyses presented here.  Finally, we present evidence that the charge distribution in the core varies strongly with the clusters and do not affect the validity of the electronic shell model. The charge variation is related to the metal-ligand bonding that also varies with the clusters. We found a covalent, ionic and polarized covalent bond for the Au$_{102}$(SR)$_{44}$, Al$_{50}$Cp*$_{12}$ and Ga$_{23}$L$_{11}$ clusters respectively.  As previously and succesfully done for the ligand-protected gold clusters, consequences into optical and charging properties, voltammetry, NMR experiments and reactivity should be expected for the both the aluminum and gallium (as well as other compostions) superatom complexes. We hope this study will lead to further theoretical and experimental investigations.    

\section*{Acknowledgments}
 We gratefully acknowledge Academy of Finland for financial support, CSC the Finnish IT Center for Science in Espoo for computational resources, and H.Gr\"onbeck and R. L. Whetten for discussions.

\section{Figures}

\begin{figure}
\centering
\includegraphics[scale=0.6]{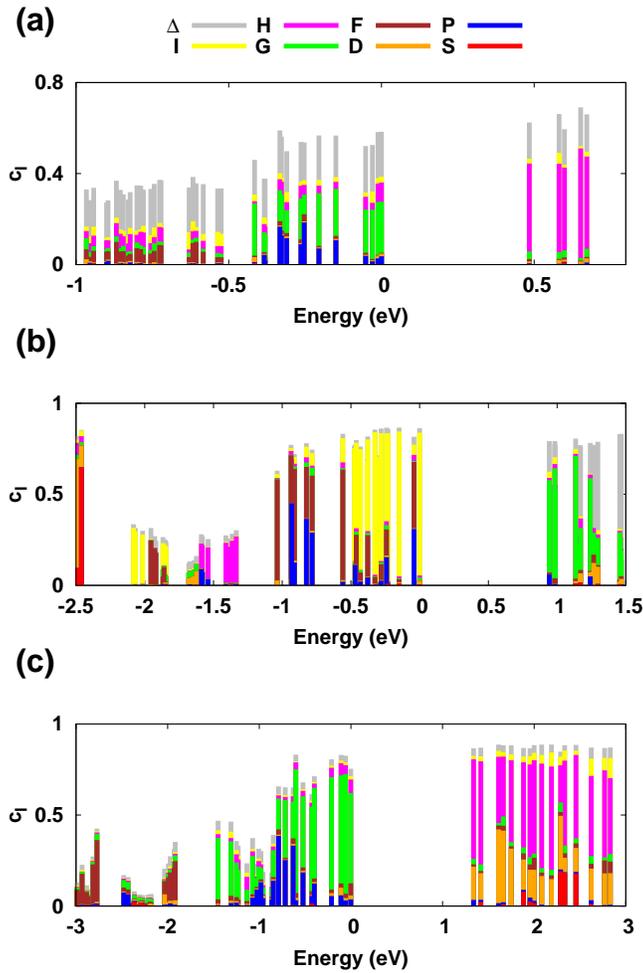}
\caption{Super-atom analysis: the angular momentum coefficient $c_l$ from Eq \ref{YLM} ($l=0,1...6$) as a function of the energy of the projected KS orbital. The difference between the sum of the coefficients  and the norm of the KS orbital inside the sphere of radius $R_0$ is $\Delta$. (a) Protected gold cluster Au$_{102}$(SC$_7$O$_2$H$_5$)$_{44}$,  the value of $R_0$ is 7.5 \AA (b) Protected aluminum cluster Al$_{50}$(C$_5$(CH$_3)_5)_{12}$ ,  the value of $R_0$ is 6 \AA (c) Protected gallium cluster Ga$_{23}$(N(Si(CH$_3)_{3}$)$_{2}$)$_{11}$,  the value of $R_0$ is 5.5 \AA. }
\label{ylm}
\end{figure}

\begin{figure}[h]
\centering
\includegraphics[scale=0.6]{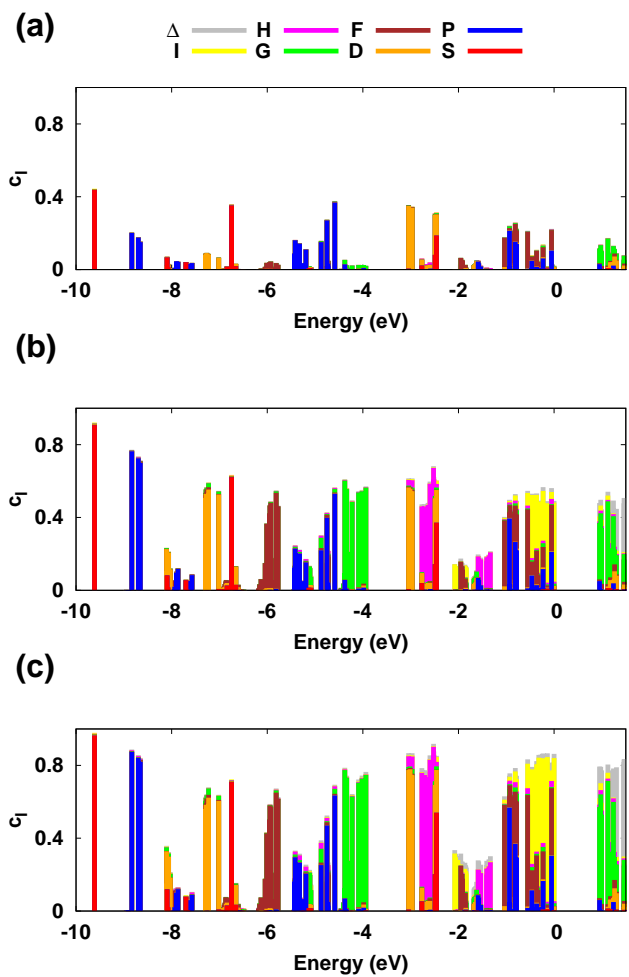}
\caption{ Supporting Information figure. Super-atom analysis: the angular momentum coefficient $c_l$ from Eq 1($l=0,1...6$) as a function of the energy of the projected KS orbital. The difference between the sum of the coefficients  and the norm of the KS orbital inside the sphere of radius $R_0$ is $\Delta$.  The analysis is shown for the protected aluminum cluster Al$_{50}$(C$_5$(CH$_3)_5)_{12}$  with $R_0$ core radius fixed at (a) 3 \AA, (b) 5 \AA and (c) 6 \AA wich corresponds to the half distance between radial layers.}
\label{ylm}
\end{figure}

\newpage

\bibliographystyle{unsrt}       % (uses file "plain.bst")
\bibliography{biblio}           % expects file "myrefs.bib"
\newpage

\end{document}